\documentclass[aps,prb,twocolumn,floats,showpacs,superscriptaddress]{revtex4-1}
\usepackage{graphicx,epsfig}
\usepackage{times}
\usepackage{graphics,dcolumn,bm,float}
\usepackage{bbold}
\usepackage{amssymb,amsmath,rotate,color}
\usepackage[pagebackref=false,colorlinks,linkcolor=blue,citecolor=blue,urlcolor=blue]{hyperref}

\newcommand{\be}{\begin{equation}}
\newcommand{\ee}{\end{equation}}

\begin{document}
\title{Topological phase diagram of the disordered 2XY model in presence of generalized Dzyaloshinskii-Moriya Interaction}
\date{\today}
\date{\today}
\author{Alireza Habibi} 
\affiliation{Department of Physics, Sharif University of Technology, Tehran 11155-9161, Iran}
\author{Rasoul Ghadimi} 
\affiliation{Department of Physics, Sharif University of Technology, Tehran 11155-9161, Iran}
\author{S. A. Jafari}
\affiliation{Department of Physics, Sharif University of Technology, Tehran 11155-9161, Iran}
   \affiliation{Center of excellence for Complex Systems and Condensed Matter (CSCM),
Sharif University of Technology, Tehran 14588-89694, Iran}
\author{S. Rouhani}
\affiliation{Department of Physics, Sharif University of Technology, Tehran 11155-9161, Iran}
\affiliation{Institute for Research in Fundamental Sciences (IPM), Tehran 19395-5531, Iran}
\begin{abstract}
Topological index of a system specifies gross features of the system. However, in situations such as strong disorder
 where by level repulsion mechanism the spectral gap is closed, the topological indices are not well-defined. 
In this paper, we show that the localization length of zero modes determined from appropriate use of transfer matrix method
reveals much more information than the topological index. The localization length can provide not only information
about the topological index of the Hamiltonian itself, but it can also provide information about the topological indices
of the "related" Hamiltonians. 
As a case study, we study a generalized XY model (2XY model) plus a generalized Dziyaloshinskii-Moriya-like (DM) interaction 
that after fermionization breaks the time-reversal invariance and is parameterized by $\phi$.
The {\em parent} Hamiltonian at $\phi=0$ which belongs to BDI class is indexed by integer winding number while the $\phi\ne 0$ {\em daughter} 
Hamiltonian which belongs to class D is specified by a $Z_2$ index $\nu=\pm 1$. 
We show that the localization length in addition to determining the $Z_2$ can count the 
number of Majorana zero modes left over at the boundary of the daughter Hamiltonian  -- which are not protected by winding number anymore.
Therefore the localization length outperforms the standard topological indices in two respects: (i) it is much faster and more accurate
to calculate and (ii) it can count the winding number of the parent Hamiltonian by looking into the edges of the daughter Hamiltonian.
\end{abstract}
\maketitle

\section{Introduction}
Band topology concerns single-particle Hamiltonians that can be classified with a topological index~\cite{SchnyderRMP2016}.
Although a complete classification has been provided~\cite{SchnyderRMP2016,Altland1997,kitaev2009periodic} but 
the consequence of the disorder and interaction are yet at scratching the surface stage~\cite{Altland2015,Gergs2016}.
For example, a $Z$-valued topological classification of non-interacting systems can be reduced to $Z_8$ classification when the interactions are taken into account~\cite{Polman2017,Fidkowski2011}. The disorder has important consequences from renormalization of parameters to a generation of 
non-translationally-invariant form of topological insulators dubbed topological Anderson insulators (TAI)~\cite{TAI2009,TAI2009Akh,TAI2012}. 
It is curious to note that sometimes the environment effect can be even helpful, as for example Bravy {\em et. al.} found 
"quenched disorder may enhance the reliability of topological qubits by reducing the mobility of Anyons at zero temperature"~\cite{bravyi2012disorder}.
 Band topology is formulated in transnational invariant systems. Although disorder breaks this important symmetry, 
disorder averaging, restores the transnational invariance, and topology become well-defined again~\cite{Prodan2015}. 
Besides, symmetry and topology, criticality  is crucial for understanding phase diagrams and the critical behavior at
Anderson transitions~\cite{Mirlin_2010}

The topological index in such systems is protected by the spectral gap, and therefore can not be changed as long as the gap is not closed~\cite{Altland2015}. 
Therefore small environmental perturbations respecting the symmetries of the topological class~\cite{SchnyderRMP2016} at hand are not
expected to change the topological index as long as they do not fill in the gap. 
These are reminiscent of the plateau-to-plateau transition in integer quantum Hall effect~\cite{Khmelnitskii,Pruisken,Pruisken2009,LevinePruisken,Joe}.
Morimoto and coworkers present a generic phase diagram of disordered topological insulators in terms of Dirac spectrum with random masses 
by studing the topology of classifying space~\cite{morimoto2015anderson}. 

A subclass of topological insulators are topological superconductors which support Majorana fermions in their spectrum.
A simple condensed matter realization of Majorana fermions is suggested by Kitaev model~\cite{kitaev2001} and its 
larger winding number generalizations~\cite{Jafari2016,Sen}.
Subsequent proposals in semiconductor-superconductor heterostructures~\cite{lutchyn2010majorana,mourik2012signatures}
has initiated a vast search for the condensed matter realizations of Majorana 
fermions~\cite{majoranadiscover1,majoranadiscover2,majoranasearch,majoranadetect}.
Studies of disorder in topological superconductors suggests that distribution of
the disorder can heavily affect the phase diagram of topological phase transitions~\cite{degottardi2013majorana}. 
Quasi-periodic systems are somehow in-between the clean and disordered systems. Models supporting Majorana fermions
in these systems such as Harper potential or Fibonacci chains are also studied. Ganeshan {\em et. al.} consider Aubry-Andr\'e-Harper model and investigate its phase diagram~\cite{ganeshan2013topological}. Cai {\em et. al.} find the boundary between topological and trivial phase for incommensurate potential~\cite{cai2013topological}. Ghadimi {\em et. al.} also consider the Fibonacci chain and they find that the topological
phase diagram itself acquires a fractal structure~\cite{ghadimi2009}. 
Given the strong dependence of topological phase diagram on deviations from periodicity (including both disorder and
quasi-periodic systems), the question that arises is, is there any genuine role played by Majorana fermions themselves in
forming the topological phase transition lines?

We have recently noticed that the localization length (LL) of Majorana zero modes not only contains  information about the topological index
which can be equally computed for clean and disordered systems, but it can also reveal information on the mechanism by which
the topological index changes from one integer value to a {\em neighboring} integer value~\cite{Habibi-resilience}. 
For this we have
taken a system in the BDI class~\cite{Altland1997} with a maximum winding number of $\pm2$~\cite{Jafari2016}, and have focused on the resilience
behavior of Majorana end modes which allows us in the hindsight to map the topological phase boundaries of the system. 
Indeed the topological phase transition is signaled by a divergence of LL~\cite{Habibi-resilience} which is in agreement
with similar results on BDI and AII  class~\cite{prodanPRL-Mondragon-Shem2014}-- both of which have time-reversal (TR) and particle-hole symmetry and are
classified with integer topological index. 
This provides us with a computationally very cheap~\cite{budich2013green,prodanPRL-Mondragon-Shem2014} and accurate
method for determination of the topological phases, which focuses on the evolution of Majorana fermions (at $E=0$) with the disorder.
The basis philosophy rests on the bulk-boundary correspondence: Simply monitor what is happenning at the edge of the system. 
This boils down to focusing on the $E=0$ (Majorana) modes which naturally live in the edges of the system.
Our larger winding number generalization of the XY model (equivalent to Kitaev model after Jordan-Wigner transformation) can 
be compared with multichannel Kitaev model which has been studied by Pekerten and coworkers who find that in the disordered system 
although gap is closed, but it is the mobility gap that decides for the change in the topological index~\cite{pekerten2017disorder}.

The emerging picture is that the disorder generates topological phase transitions between gapped
states~\cite{Altland2015}, the boundary of which is marked by {\em zero energy extended states}~\cite{Habibi-resilience,prodanPRL-Mondragon-Shem2014}.
Now imagine that a TR breaking agent is introduced. This TR breaking agent will work against localization and tries to
create extended states~\cite{Furusaki_1999}. The ensuing extended states are expected to intervene with the topological phase transitions that
are driven by the disorder. For this purpose we further extend the 2XY model which belongs to BDI class with a similar
extension of Dzyaloshinskii-Moriya~\cite{Dzyaloshinskii,Moriya} (DM) interaction -- the clean limit of which still remains solvable -- that is parameterized
by the TR breaking parameter $\phi\ne 0$. We use the terminology of the parent (daughter) Hamiltonian to refer to the $\phi=0$ ($\phi\ne 0$) 
Hamiltonian which belongs to BDI (D) class and is classified by a $Z$ ($Z_2$) index. 
As far as the $Z_2$ topological index is concerned, all phases with even number of pairs of Majorana end modes
are equivalent to zero pairs of Majorana fermions. Therefore the 
ensuing $Z_2$ index of the daughter Hamiltonian is not able to distinguish between the even numbers of
pairs of Majorana end modes -- which are inherited from the $\phi=0$ parent Hamiltonian. 
Similarly, all phases with odd number of Majorana end modes -- as far as the $Z_2$ index
is concerned -- are equivalent to one pair of Majorana end modes. 
The localization length of the zero-energy states
signals the disorder threshold at which pairs of Majorana fermions are drawn into the bulk of
Anderson localized states in a one-by-one fashion~\cite{Habibi-resilience}. This picture
persists in the daughter Hamiltonian with $\phi\ne 0$ where the topological index is not 
even a winding number anymore. Therefore by looking into the edge modes of the daughter
Hamiltonian, the LL is able to tell us about the number of Majorana zero modes left from
the parent Hamiltonian. 

The roadmap of the paper is as follows: In section~\ref{model.sec} we introduce a generalization
of the DM interaction in the spirit of nXY model~\cite{Jafari2016} which allows for exact Jordan-Wigner solvability
in the clean limit. In section~\ref{footprint.sec} we discuss how the footprints of the winding number of the parent Hamiltonian
survive in the daughter Hamiltonian whose TR breaking parameter $\phi$ is non-zero. In section~\ref{phasediag.sec} we discuss the
phase diagram of the model and compare the standard Pfaffian and localization length diagnosis tools. 
We end the paper with our conclusion and outlook.

\section{Model and Method}
\label{model.sec}
In this section, we will extend the model previously introduced by one of the authors~\cite{Jafari2016} that allows to
engineer arbitrarily large winding numbers. We consider the consequence of disorder and time reversal breaking term on the phase diagram
of this model. Consider 2XY model extended by DM and disorder terms as follows:
\begin{equation}
    H=H^{\textrm{2XY}} +H^{\textrm{DM}}+H^{\textrm{dis}}.
\end{equation}
    $H^{\textrm{DM}}$ is a generalization of Dzyaloshinskii-Moriya interaction, $H^{\rm dis}$ is random transverse field term
which are defined as follows: 
\begin{widetext}
\begin{subequations}
\begin{align}
H^{\textrm{2XY}} &=
    \sum_j{(J'_1 +\lambda_1) \sigma_j^x\sigma_{j+1}^x
    +{(J'_1-\lambda_1)\sigma_j^y\sigma_{j+1}^y
    +(J'_2 +\lambda_2)\sigma_j^x\sigma_{j+1}^z \sigma_{j+2}^x} 
    +(J'_2-\lambda_2)\sigma_j^y\sigma_{j+1}^z\sigma_{j+2}^y  } ,
\\
H^{\textrm{DM}}&= 
\sum_j{\Delta_1 (\sigma_j^x\sigma_{j+1}^y-\sigma_j^y\sigma_{j+1}^x)} 
+{\Delta_2 (\sigma_j^x\sigma_{j+1}^z\sigma_{j+2}^y-\sigma_j^y\sigma_{j+1}^z\sigma_{j+2}^x)},
\\
H^{\textrm{dis}}&=\sum_j{\left(\varepsilon_j +\mu\right)\sigma_j^z}.
\end{align}
\end{subequations}
\end{widetext}
By using  Jordan Wigner transformation,
    \begin{align}
    \sigma_j^z=2c_j^\dagger c_j -1,
    \sigma_j^x=e^{i \phi_j} (c_j^\dagger+ c_j),
    \sigma_j^y=i e^{i \phi_j} (c_j^\dagger- c_j),
    \end{align}
    the Hamiltonian can be translated to fermionic language.  $ \phi_j=\pi \sum_{l<j}{c_l^\dagger c_l} $ is the phase string, which 
serves to guarantee the anti-commutative requirement of fermions. So by neglecting constant terms,  fermionic version of this model given by: 
\begin{multline}
H=2\sum_{j} \sum_{s=1,2}{J_s e^{i s \phi} c^\dagger_j c_{j+s}+\lambda_s c_j^\dagger c_{j+s}^\dagger} + h.c\\
+2\sum_{j}{\left(\mu+\varepsilon_j\right) (c^\dagger_j c_{j}-\frac{1}{2})} 
\label{extendedkitaev}
\end{multline}
where the following re-parameterization is introduced,
 %   \begin{gather}
         \begin{align*}
\begin{matrix}
     J'_1 &\rightarrow J_1 \cos{\phi} ,&\quad&
     J'_2 &\rightarrow J_2 \cos{2\phi} ,&\quad&
     \Delta_1 &\rightarrow J_1 \sin{\phi} \\
     \Delta_2 &\rightarrow J_2 \sin{2 \phi}.&\quad&
%     J_0^j &\rightarrow \varepsilon_j +\mu.
\end{matrix}
 \end{align*}
 %   \end{gather}
   This extension of the Kitaev model not only extends the hoppings and pairing to longer ranges, but also 
includes appropriate extension of the DM interaction which is encoded in the parameter $\phi$. 
The $H^{\rm 2XY}$ portion of the above Hamiltonian belongs to the BDI class~\cite{SchnyderRMP2016} and allows for
integer topological index which in this case turn out to be between $-2$ and $2$~\cite{Jafari2016}.
Adding the $H^{\rm dis}$ still preserves the winding number for small disorder strength. However, increasing the disorder causes
the winding number to be lost one-by-one~\cite{Habibi-resilience}. The bulk-boundary correspondence allows to detect the
reduction of winding number by only focusing on the boundary (Majorana) modes that are locked to $E=0$ energy~\cite{Habibi-resilience}.
Breaking the TR by smallest amount of $\phi$, reduces the BDI class to the class D which admits a $Z_2$ topological index.
The standard method to calculate the $Z_2$ index is to compute the Pfaffian~\cite{Bagrets_2012,kitaev2001,Ardonne_pf}. 
Note that the above generalized DM interaction breaks the TR, but it is not equivalent to an applied magnetic field. 
It only couples to the hopping term on Jordan-Wigner fermions, but does not couple to the condensate. 
Therefore there is no Meissner effect associated with TR breaking arising from non-zero $\phi$.
    
\subsection{Longer range Kitaev chain}
Here we wants to study the extended 1D Kitaev model with onsite disorder:
    \begin{multline}
H=2\sum_{j} \sum_{s=1,r}{J_s c^\dagger_j c_{j+s}+\lambda_s c_j^\dagger c_{j+s}^\dagger}+ h.c\\
+2\sum_{j}{\left(\mu+\varepsilon_j\right) (c^\dagger_j c_{j}-\frac{1}{2})}
\end{multline}
Although in this work we are interested in the $r=2$ case, but even for a general $r$ this Hamiltonian belongs to BDI class
where the topological index is an integer $n_w$ which for this model satisfies $ |n_w| \le r $.
By breaking  TR symmetry, the system will belong to class D, which in one dimension admits a $Z_2$ topological index.
Let us see in detail how this happens. 
The Jordan-Wigner fermionized version of the model will be,
\begin{multline}
H=2\sum_{j} \sum_{s=1,r}{J_s e^{i s \phi} c^\dagger_j c_{j+s}+\lambda_s c_j^\dagger c_{j+s}^\dagger}+ h.c\\
+2\sum_{j}{\left(\mu+\varepsilon_j\right) (c^\dagger_j c_{j}-\frac{1}{2})} 
\end{multline}
This Hamiltonian can also be represented in terms of the Majorana fermions, $ a_j=(c_j^\dagger+c_j)$
and $b_j=i (c_j^\dagger-c_j)$ which satisfy the fallowing algebra:
\begin{gather}
\begin{matrix}
     \{a_i,a_j\}=\{b_i,b_j\}=2\delta_{i,j},\\
 a_i^\dagger =a_i \text{ and }  b_i^\dagger =b_i.
\end{matrix}
\end{gather}
In high energy physics, the particle that satisfy mentioned algebra are called Majorana fermions which are their own antiparticles. 
The Majorana representation of our Hamiltonian becomes, 
\begin{widetext}
\begin{align}
H&=
i\sum_{j} \sum_{s=1,r} \bigg(
\left( J_s  \cos{s \phi}-\lambda_s\right) a_j b_{j+s} 
+\left( -J_s  \cos{s \phi}-\lambda_s\right)b_j a_{j+s} +J_s  \sin{s \phi} \; a_j a_{j+s}
+  J_s  \sin{s \phi} \; b_j b_{j+s}
\bigg)
\nonumber \\
&+i\sum_{j}\bigg( \frac{\mu+\varepsilon_j}{2}\; a_j b_j
-\frac{\mu+\varepsilon_j}{2} \; b_j a_j \bigg)
\end{align}
For a clean system, using Fourier transformation it can be rewritten as, 
\begin{gather}
H=\sum_k{\psi_k^T h(k) \psi_{-k}}\qquad, 
\psi_k^T=\begin{pmatrix}
a_k & b_k 
\end{pmatrix},\\
\frac{h(k)}{2}=\left(\sum_{s=1,r} {J_s  \sin{sk} \sin{s\phi} } \right)\tau_0 
+    \left(\frac{\mu}{2}+\sum_{s=1,r} {J_s \cos{sk} \cos{s\phi}}\right)  \tau_y
+\left(\sum_{s=1,r} {\lambda_s \sin{sk}} \right)\tau_x.
\label{hk_clean}
\end{gather}
In this basis TR can be represented as $\tau_z K$. The action of this operator is $a_k \to a_{-k}$ and $b_k \to -b_{-k}$. Note that under TR $h(k) \to \tau_z K h(-k) \tau_z K $. This operation leaves second and third term invariant while the first term arising from the non-zero $\phi$ is not invariant. Furthermore, PH in this basis can be represented as PH$=\tau_0 K$ which acts as $a_k \to K a_k=a_{-k}K$ and $b_k \to K b_k=b_{-k}K$, such that PH$^2=+1$. The effect of PH on the matrix $h(k)$ is therefore given by $\tau_0K h(-k)\tau_0K$. 
Another way to see why the PH is identified as above is as follows: First of all, since PH is an anti-unitary operation, it must involve
the complex conjugation operator $K$. 
To make the first term change sign, it should also involve $k\to -k$. 
As for the matrix part, as far as the first term of the above Hamiltonian is concerned, it can be any of $\tau_\mu$ with $\mu=0,1,2,3$. 
So the first term does not constraint the matrix part. Now let us move to the second term. The operation $k\to -k$ does not do anything
to $\cos sk$. The complex conjugation, however, produces a minus sign as it acts on $\tau_y$. Since the required minus sign is already
produced, the matrix part must commute with $\tau_y$. This leaves us only with two choices, namely $\tau_0$ or $\tau_y$. 
The third ($\tau_x$) term completely fixes this. The $\sin sk$ gives a minus sign under $k\to -k$, and $\tau_x$ does not care
about complex conjugation $K$. Therefore the matrix part must also commute with $\tau_x$. This fixes the matrix part of PH to be $\tau_0$. 
Now, we define ${\cal C}=$TR.PH$=\tau_z$. The above Hamiltonian lacks ${\cal C}$ and TR but possesses the above PH symmetry that squares to $+1$. Therefore, our Hamiltonian belongs to class D hence in 1D admits a $Z_2$ classifications\cite{SchnyderRMP2016}. 
Energy spectrum can be
find by diagonalizing Hamiltonian:
\begin{gather}
\label{exact-energy}
\frac{E(k)}{2}=\left(\sum_{s=1,r} {J_s\sin{sk}\sin{s\phi} } \right)
\pm \sqrt{\left(\frac{\mu}{2}+ \sum_{s=1,r} {J_s\cos{sk} \cos{s\phi} }\right)^2+\left(\sum_{s=1,r}{\lambda_s\sin{sk}  }\right)^2}
\end{gather}
\end{widetext}
As can be seen, under the sequence of operations, $k\to -k$ and acting on matrix part from both sides by $\tau_0K$, the 
eigenstates $|k,\pm\rangle$ is mapped to $-|k,\mp\rangle$. This transformation, 
sends a state with energy $E$, to a state with energy $-E$ which is
the manifestation of PH symmetry on the spectrum. 

\section{Footprints of winding number in the $Z_2$ phase}
\label{footprint.sec}
\begin{figure}
    \centering
    \includegraphics[width=1\linewidth]{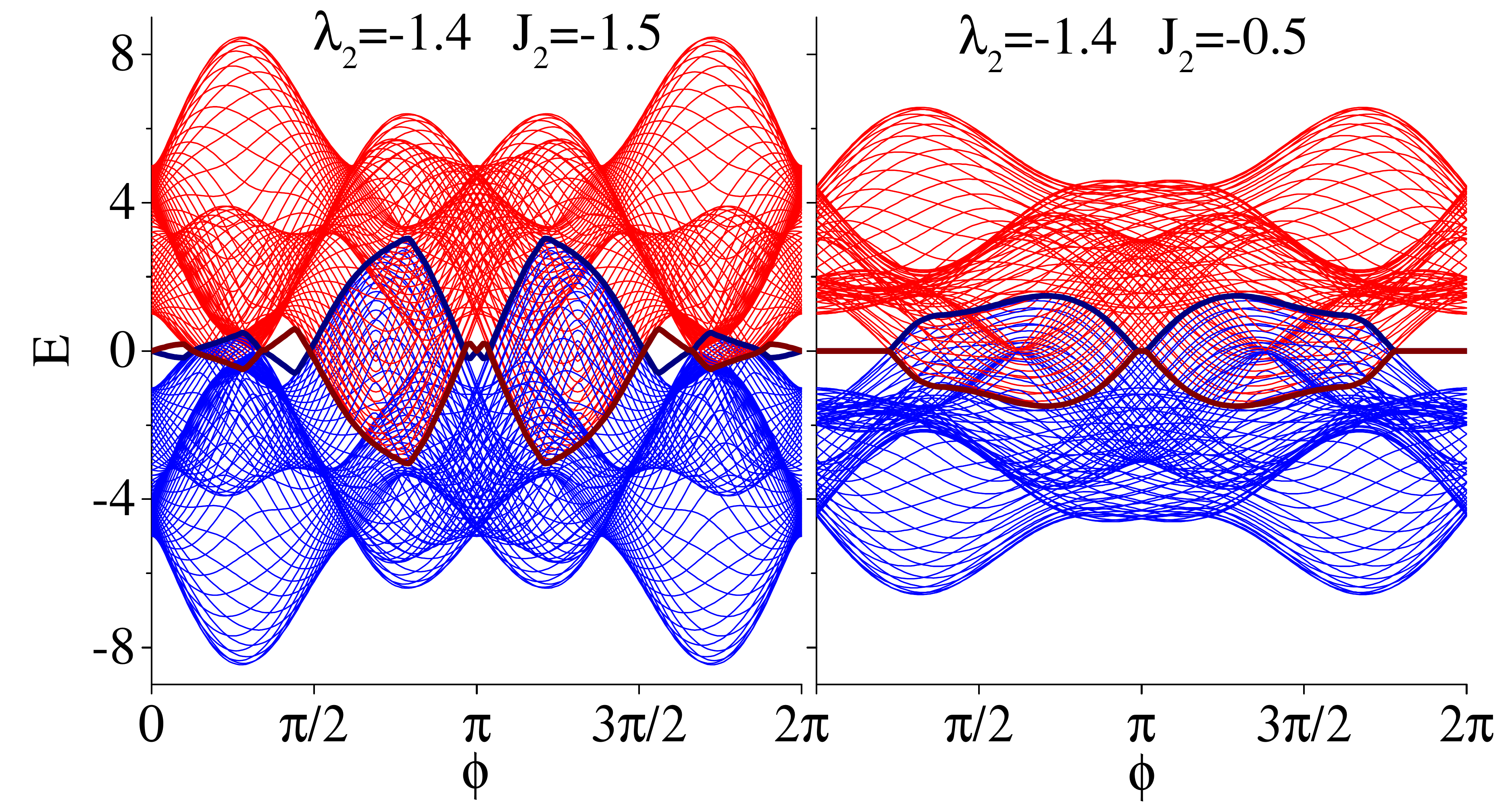}
    \caption{(Color online) Tower of states of the clean limit of Eq.~\eqref{extendedkitaev} with open boundary conditions as a function 
of the generalized DM parameter $\phi$. This spectrum drawn for $\mu=0$ and $J_1=\lambda_1=1$ for a system of $L=10^3$ sites. 
For clarity the continuum part of bands has been extracted from Eq.~\eqref{exact-energy}. 
(left): For $\lambda_2=-1.4$ and $J_2=-1.5$ there are $2|n_w|=4$ mid-gap states and the system has trivial $Z_2$ index, $\nu=1$. 
(right): $\lambda_2=-1.4$ and $J_2=-0.5$ and $2|n_w|=2$ mid-gap states and the system has non-trivial $Z_2$ index, $\nu=-1$. 
Red (blue) lines show top (bottom) band. Dark red (navy blue) line shows the mid-gap States. 
For those values of the generalized DM parameter $\phi$ where there is a clear separation between top and bottom bands, 
the Majorana fermions of the parent Hamiltonian continue to leave in the mid-gap. When the top and bottom bands merge, 
the mid-gap states are drowning into continuum. 
}
    \label{fig:energy-band}
\end{figure}

In Fig.~\ref{fig:energy-band}  we plot the tower of states for the clean limit of the Hamiltonian~\eqref{extendedkitaev}
for $\mu=0$, $\lambda_2=-1.4$ and $J_2=-1.5$ for a continuum of generalized DM parameter $\phi$. Red (blue) lines indicate the top (bottom) energy states.  
The role of generalized DM parameter $\phi$ is to reduce the Hamiltonian from class BDI ($Z$ classification) to class D ($Z_2$ classification). 
The $\phi=0$ parent model is classified with an integer winding number, $n_w$. The left panel 
corresponding to parameters $\lambda_2=-1.4, J_2=-1.5$ (note that we always take $\lambda_1=J_1$ where $J_1$ is unit of energy)
which is specified with a winding number of $-2$~\cite{Jafari2016}, while the panel in the parent model is specified by the winding number of $-1$. This means that for the parent Hamiltonian at $\phi=0$, in the left panel there are
two ($|n_w|=2$) pairs of zero modes (i.e. four zero energy states) while in the right panel there are only one ($|n_w|=1$) pair of 
zero modes (i.e. two zero energy states)~\cite{Habibi-resilience}. For a representative point in the
phase diagram of the $\phi=0$ model~\cite{Jafari2016} that corresponds to $n_w=0$ there are no mid-gap states whatsoever. 

Any non-zero value of the generalized DM parameter $\phi$ reduces the BDI class to the D class, 
and hence the non-zero $\phi$ model must be classified with a $Z_2$ invariant, $\nu=\pm1$.
The $\nu=+1 (-1)$ corresponds to trivial (nontrivial) topology. 
As far as the $Z_2$ index for $\phi\ne 0$ model is concerned, all even (odd) winding numbers
of the parent Hamiltonian correspond to $\nu=1 (-1)$ or more compactly 
\be
\nu=(-1)^{n_w}.
\label{nunw.eqn}
\ee
As far as the above topological index is concerned, for the $\phi\ne 0$ daughter Hamiltonian, 
all the even winding numbers of the parent ($\phi=0$) Hamiltonian are the same. But as can be seen
in Fig.~\ref{fig:energy-band}, the winding number of the parent Hamiltonian still have its footprints in the $\phi\ne 0$ model. For every value of $\phi$ where there is a clear separation between the positive and negative energy states, there are $2|n_w|$ mid-gap states. Being
a mid-gap state, they are localized in the edge. The mid-gap states in Fig.~\ref{fig:energy-band}-left
are not protected as they correspond to the trivial $Z_2$ topological index, $\nu=+1$ and hence dissever 
from $E=0$ for nonzero values of the generalized DM parameter $\phi$. In contrast, the mid-gap states 
in Fig.~\ref{fig:energy-band}-right remain pinned to $E=0$ as they are protected by the $Z_2$
topological index, $\nu=-1$. Note that for those values of the generalized DM parameter $\phi$ that positive and
negative energy bands merge, the $Z_2$ index can not even be defined.

The mid-gap states of the left panel
used to be Majorana zero modes in the $\phi=0$ parent Hamiltonian, but in the $\phi\ne 0$ daughter Hamiltonian
they are not Majorana zero modes anymore. However, they are still separated from the rest of the spectrum
and are hence localized in the edge. These mid-gap states are not as localized as the $\phi=0$, as the non-zero 
$\phi$ itself is increasing the localization length of the mid-gap states. 
By increasing $\phi$  they eventually enter the continuum of extended states.  This is how the TR breaking
agent $\phi$ causes the delocalization of mid-gap states. 
For odd values of the winding number such as those in Fig.~\ref{fig:energy-band}-right,
Majorana zero modes of the parent Hamiltonian at every edge, hybridize in a pairwise fashion and hence get dissevered from $E=0$, 
leaving behind one zero mode which still remains protected. For the even values of the winding number such as those in Fig.~\ref{fig:energy-band}-left
all zero modes at a given edge hybridize with each other, and are therefore pushed away from $E=0$, but they still
remain localized in the edge, although not topologically protected. 
As long as we are dealing with a clean system, the edge modes are not harmed by the rest of
the spectrum. Therefore, although the daughter Hamiltonian is characterized by $\nu$, it still remembers the $n_w$ of the parent BDI Hamiltonian by having $|n_w|$ mid-gap states at every edge. 

\begin{figure}[b]
    \centering
    \includegraphics[width=0.8\linewidth]{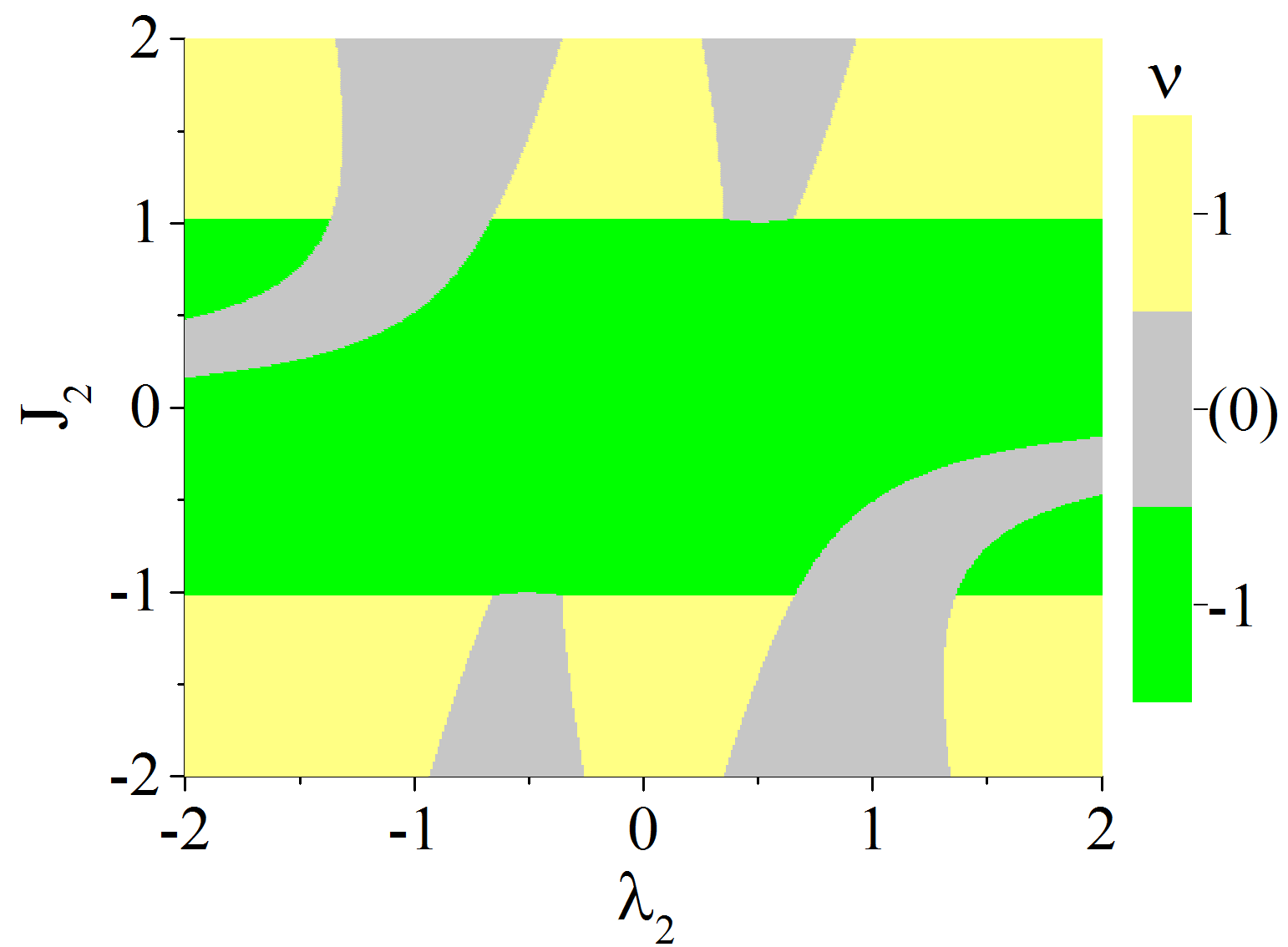}
    \caption{(Color online) Phase diagram of the 2XY model with generalized DM interaction $\phi=0.1$. The phase diagram 
    consists in regions with $\nu=\pm 1$ which are spawn from regions with winding number $n_w$ of the $\phi=0$
    model and satisfy $\nu=(-1)^{n_w}$. The gray region corresponds to gapless region. This region only separates
    between regions that have the same $\nu$. Regions with opposite $\nu$ are separated by a gapless {\em line}. 
    }
    \label{fig:jlamdaphi}
\end{figure}
\section{Phase diagram of clean $Z_2$ system}
\label{phasediag.sec}
Upon deviation of the generalized DM parameter $\phi$ from zero, the topological index
of the system will be a $Z_2$ invariant, and the winding number $n_w$ of the parent system will be remembered as a number of (not all-protected) mid-gap states. 
The phase diagram of the parent 2XY Hamiltonian consists of regions with the definite
winding number $n_w$ which are separated by gapless lines~\cite{Jafari2016,Habibi-resilience}. 
The boundaries of the parent Hamiltonian are of two types: (i) gapless lines across which the
winding number changes by one and (ii) gapless lines across which the winding number changes by two. Now the question is, what happens to the phase diagram as the generalized DM parameter $\phi\ne 0$
is introduced?

First of all as in Eq.~\eqref{nunw.eqn} all regions of the parent Hamiltonian that have even (odd) $n_w$, in the
daughter Hamiltonian will correspond to trivial (non-trivial) $Z_2$ index, $\nu=+1 (-1)$. However, those borders that
separate same $\nu$ index, will be broadened to a gapless region, rather than a gapless line. This is depicted in 
Fig.~\ref{fig:jlamdaphi}. The gray region in this figure denotes the gapless region which always separates two 
regions having the same $\nu$. In the language of the parent Hamiltonian, this gapless region always
occurs between regions of that have the same winding number parity. 

Let us see how does the broadening of the gapless line into a gapless (gray) region happens. 
In order to have a gapless region in parameter space, 
there should exist a $ 0 \leq k \leq 2\pi $ such that $ E^2(k)=0 $. Using Eq.~\eqref{exact-energy} for the most general case $r=N$
we have,
\begin{widetext}
\begin{gather}
\sum_{s=0}^N\sum_{s'=0}^N
\left[\lambda_s \lambda_{s'} +J_s J_{s'}  \cos{(s+s')\phi}\right]\cos{(s-s')k} 
+    \left[-\lambda_s \lambda_{s'}  +J_s J_{s'}  \cos{(s-s')\phi}\right]\cos{(s+s')k}=0,
\end{gather}
where $ J_0=\mu/2 $ and $ \lambda_0=0 $.  For  $ r=2 $ this equation reduces to,
\begin{gather}
\left[\lambda_1 \lambda_2+J_1 J_2 \cos{3\phi} + \mu J_1 \cos{\phi}\right] \cos {k}
+\frac{J_1^2-\lambda_1^2+2 \mu J_2 \cos{2\phi}}{2} \cos {2k}
+\left[-\lambda_1 \lambda_2+J_1 J_2 \cos{\phi}\right] \cos {3k}  \nonumber\\
+\frac{J_2^2-\lambda_2^2}{2} \cos {4k} 
+\dfrac{\lambda_1^2+\lambda_2^2+J_1^2 \cos{2 \phi}+J_2^2 \cos{4 \phi}}{2} 
+\dfrac{\mu^2}{4}
 =0
 \label{ep2r2}
\end{gather}
\end{widetext}
 Two possible gapless points are $k = 0 , \pi$ (note that $-\pi$ is equivalent to $+\pi$) which result in,
\be 
 J_2 =\left(-\xi J_1 \cos{\phi}- \mu/2 \right)/\cos{2\phi},
 \label{borderpm1.eqn}
\ee
where $\xi=\pm 1$ corresponds to $\cos{k}$ with $k=0,\pi$. 
Already at $\phi=0$ we obtain the horizontal borders $J_2=\mp J_1$ of the 2XY model~\cite{Jafari2016,Habibi-resilience}.
As for the other border, let us start with the zeroth order border corresponding to $\phi=0$ which 
corresponds to a gap-closing at $ k=\arccos{(-\lambda_1/{2\lambda_2})}$ and gives the border line,~\cite{Jafari2016,Habibi-resilience},
\begin{align}
  J_2 =\frac{\dfrac{J_1\lambda_1}{2\lambda_2}-\frac{\mu}{2}}{\frac{\lambda_1^2}{2\lambda_2^2}-1}  \qquad \text{ for }  \!  \left|\dfrac{\lambda_1}{2\lambda_2}\right|\leq 1
\end{align}
Now let us Taylor expand for small $\phi$ to see how does the gapless line broadens into a gapless region 
upon turning a very small generalized DM parameter $\phi$. 
The gray gapless region in the absence of disorder will be a metallic region. 

Now let us turn on the calculation of the $Z_2$ topological index $\nu$ for the 
$\phi\ne 0$ Hamiltonian. Our Hamiltonian Eq.~\eqref{hk_clean} is of the following 
generic form,
\begin{equation*}
H=\sum_k \psi_k^\dagger (d_0\sigma+\vec{d_k}.\vec\sigma) \psi_k\qquad \psi_k=\left(\begin{matrix}
a_k\\b_k
\end{matrix}\right)
\end{equation*}
In our case the $k$-even component is given by $d_y(k)=\sum_{s=1,r}J_s\cos(sk)\cos(s\phi)+\mu/2$ 
which hence characterizes the $Z_2$ index~\cite{Alicea}
\begin{equation}
\nu=sign(d_y(0)d_y(\pi))
\label{nusign.eqn}
\end{equation}
For the above $d_y(k)$ function it becomes,
\begin{multline}
\nu=sign\left(\frac{\mu}{2}+\sum_{s=1,r} {J_s  \cos{s\phi} }\right)\times \\
sign\left(\frac{\mu}{2}+\sum_{s=1,r} {J_s \cos{s\pi}  \cos{s\phi}}\right) 
\end{multline}
It is important to note that the above formula works as long as the system is gapped. 
It can not be applied to the gray region in Fig.~\ref{fig:jlamdaphi}. Outside this region, it is consistent with this figure. 

Indeed in the Majorana representation, the Hamiltonian can be represented by a skew matrix $h$ as,
\begin{equation*}
H=i \sum \gamma_{i}h_{ij}\gamma_j
%\qquad \gamma_i=\begin{matrix}
%i=2n&&a_n\\
%i=2n+1&&b_{b}
%\end{matrix}
\end{equation*}
where $h_{ij}$ are matrix elements of $h$ and $\gamma_i$ are Majorana fermion creation operators. 
The $Z_2$ index is given by the Pfaffian~\cite{Ardonne_pf},
\begin{align}
\nu=\text{sign \!}\text{Pf}\left(H\right)
\end{align}
Using the above definition and the Wimmer package for calculation of  Pfaffian~\cite{Wimmer2012},
the phase diagram of Fig.~\ref{fig:jlamdaphi} can be produced. In the gapped region it agrees with 
formula~\eqref{nusign.eqn}. 
For the clean system, increasing the size of system do not change the \text{sign} of  Pfaffian gaped phases.  
On the opposite side, in the gapless phase sign of Pfaffian changes by changing the length of the system somehow randomly. 
Therefore in the gapless (gray) region of Fig.~\ref{fig:jlamdaphi}, the numerical calculation of Pf
gives a strongly fluctuating Pfaffian sign. The random fluctuations are controlled by size (in the clean system) and/or disorder. 
This is simply because for a gapless system the Pfaffian can not be defined. In this case, however, this strong sign fluctuations can serve as a convenient tool to determine the broadening of the gapless region for $\phi\ne 0$ system. 

To summarize this section, the generalized DM parameter $\phi$ broadens the gapless lines of the $\phi=0$ model to gapless regions that separate regions with the same $\nu$. In the case of regions with different $\nu$, 
the generalized DM parameter $\phi$ only shifts the gapless line that separates them without any broadening. 
The gapless region in numerical calculations is signaled as strong fluctuations in the sign of Pfaffian. 
As long as there is no disorder, this gapless region is metallic. Now we wish to study
what happens when we turn on the disorder.

\section{Z2 topological phases with onsite disorder}
In one dimension, the smallest amount of uncorrelated on-site disorder localize wave functions and makes the systems Anderson insulator.
Anderson insulator is distinct from band insulator in that despite a gapless spectrum, the conduction ceases because of the localization of wave-functions. 
In such a situation, the concept of localization length is used to quantify the localization of wave functions.  
Localization length indicates, how much a given wave function is localized which generally depends on energy and disorder strength. 

Our previous study reveals that in the parent BDI Hamiltonian, the localization length of $E=0$ modes is capable of
sharply identifying the onset of disorder strength at which the winding number changes. At this threshold values, 
one pair of Majorana fermions across the two ends of the system become critically delocalized which allows them to hybridize and are therefore drown into the bulk of Anderson localized states~\cite{Habibi-resilience}. 
In the hindsight, the localization length is able to assign the winding number to each phase. 
The location of divergence of localization length of the $E=0$ wave functions identifies the phase boundaries of the parent BDI 
Hamiltonian~\cite{Habibi-resilience}. Moreover, the calculation of localization length can be efficiently and precisely performed with an appropriate modification of the transfer matrix method (see appendix~\ref{tm.sec}). 
This observation elevates the localization length as extracted from transfer matrix method to a diagnosis 
tool that can reveal information about the topology (winding number) of the system. For details please
see Ref.~\onlinecite{Habibi-resilience}. 
%\begin{figure}
%    \centering
%    \includegraphics[width=1\linewidth]{./l-j-mu-w10}
%    \caption{Localization length and winding number($Z$ topological phase) for different value of constant chemical potential strength plus a constant noise in it given by $\mu\pm w$, which "noise" set to be $w=10$. it is clear that constant chemical potential asymmetries the phase diagram, and over-all system need more coupling to being topological. Localization length show band touching in some region, which may be topological phase transition or simple band closing.}
%    \label{fig:l-j-mu-w}
%\end{figure}
%Here we extend this analysis to including chemical potential which shown in Fig. \ref{fig:l-j-mu-w}.
%As we can expect, the topological phase diagram has been changed with modulating disorder and chemical potential.
%As we can see for specific topological phase, increasing chemical potential or disorder needs more pairing and hopping interaction.

\begin{figure}
    \centering
    \includegraphics[width=\linewidth]{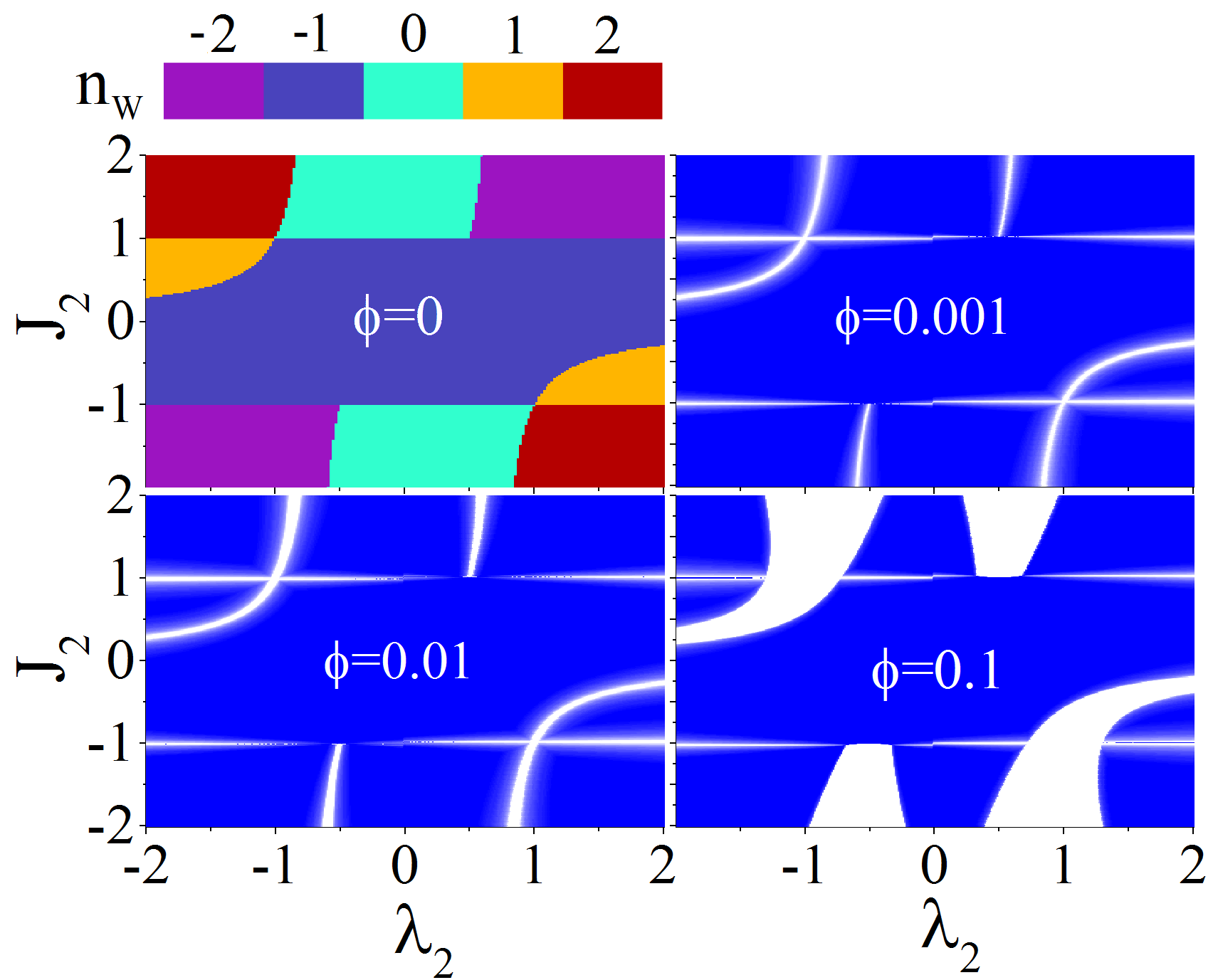} \label{fig:mu04}
    \caption{(Color online) Localization length for different value of $\phi $ in the clean system: 
    this figure shows the grow of gapless phase with extended wave functions by increasing generalized DM parameter $\phi$. 
    As we can see the broadening of transition lines are very sensitive to the strength of the $\phi$ perturbation. 
    Note that in agreement with Fig.~\ref{fig:jlamdaphi}, only borders separating the same $Z_2$ topological index $\nu$ 
    (separating phases with only even, or only odd winding numbers in parent Hamiltonian)
    will be broadened by increasing $\phi$. This figure corresponds to $J_1=\lambda_1=1$ and $\mu=0$.}
    \label{fig:/W00phi}
\end{figure}

Upon turning on the generalized DM parameter $\phi$, the topological index of the daughter Hamiltonian will not be an integer (winding) number anymore. But it will turn out that the localization length of $E=0$ states of the $\phi\ne 0$
the model will still remember information about the winding number of the parent Hamiltonian. 
%Due to gap closing, this definition breaks and the system goes to non-topological states such as metal or Anderson insulator. \\
For an extension of Kitaev model without disorder, the role of time-reversal symmetry breaking by the parameter $\phi$ 
is to broaden the gapless topological phase transition boundaries of the $\phi=0$ model into a gapless region~\cite{Sen}. 
In our model, we have used the localization length of $E=0$ modes to produce the phase diagram of the clean system in Fig.~\ref{fig:/W00phi}. 
The top-left panel shows the phase diagram of the parent $\phi=0$ model with sharp boundaries that separate regions with
various winding numbers. Across the two horizontal boundaries at $J_2=\pm 1$ the winding numbers differ by one.
Therefore they separate even winding numbers from odd winding numbers. Other phase boundaries separate regions
across which the winding number changes by two. In agreement with Fig.~\ref{fig:jlamdaphi}, only the later are
broadened into a gapless region, while the former are slightly shifted in agreement with Eq.~\eqref{borderpm1.eqn}. 
As emphasized in Fig.~\ref{fig:jlamdaphi}, the gapped phases separated by broad regions corresponds to the same
$Z_2$ index. 

\begin{figure}[t]
    \centering
        \includegraphics[width=\linewidth]{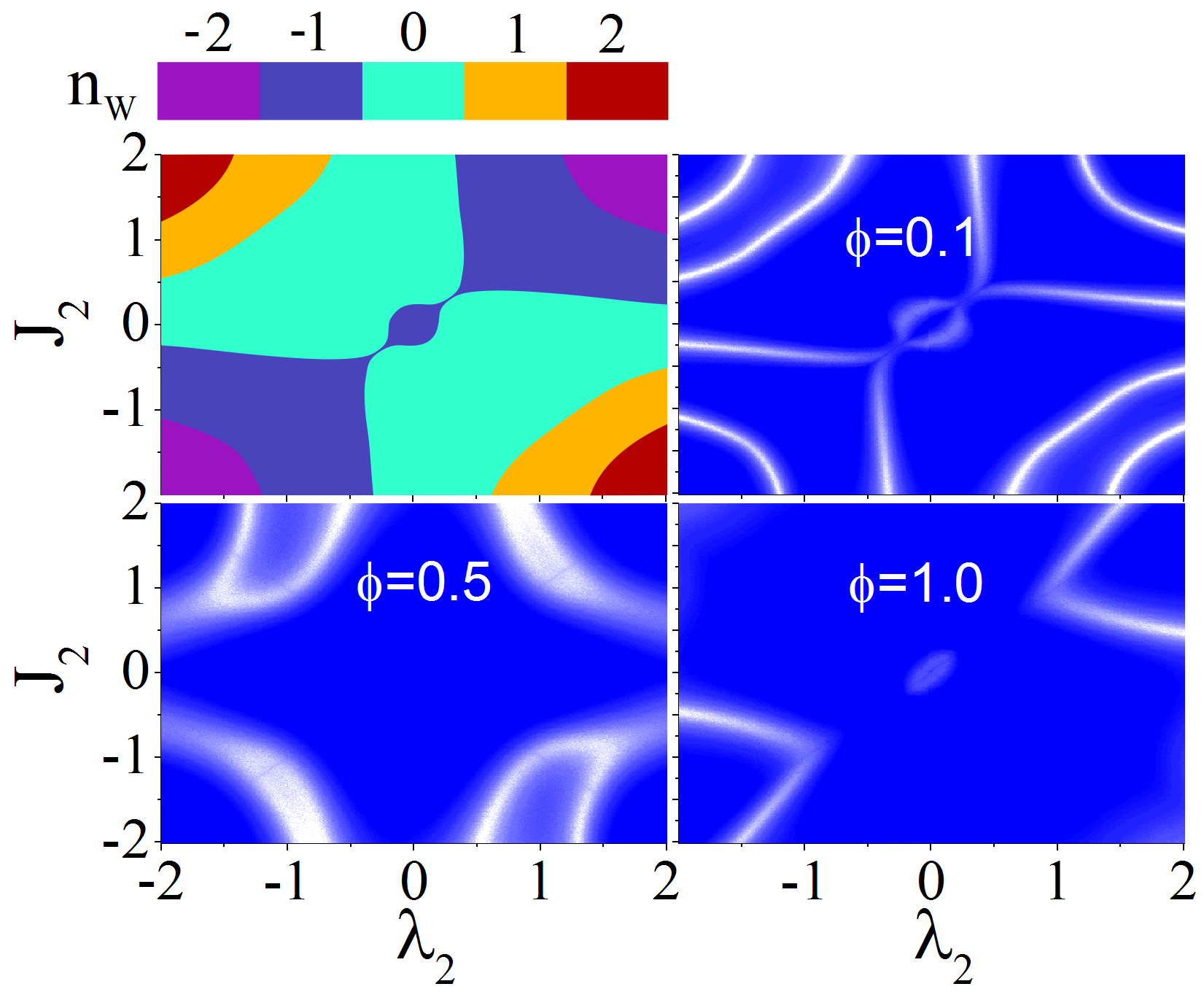} \label{fig:mu08}
    \caption{(Color online) Localization length for different value of $\phi $ and disordered strength give by $ W=10 $.
    The role of $\phi$ is to broaden the bright region separating the same $\nu$ phases, while the
    disorder works in opposite direction, as it leads to Anderson localization of the $E=0$ states. The winding number
    map in the top left panel is obtained from the LL itself~\cite{Habibi-resilience}. 
    }
    \label{fig:jlambdaforphi}
\end{figure}
Now let us study the interplay between the $\phi$ perturbations and disorder. 
As pointed out, starting from $\phi=0$ limit, the role of $\phi\ne 0$ is to broaden the border between
the same $\nu$ phases. Comparing the same values of $\phi$ in Fig.~\ref{fig:jlambdaforphi} (corresponding to disorder strength
$W=10$) with the corresponding part of Fig.~\ref{fig:/W00phi} of the clean system, one sees that the
broadening of the gapless region is reduced by the disorder. 

Despite that for the $\phi\ne0$ Hamiltonians the winding number can not be defined, 
however, still the divergence of the localization length on the white borders signals
something happening in the spectrum. This is nothing but the remnants of the integer
winding number genome of the parent Hamiltonian which shows up as enhancements of the
localization length of the zero modes and therefore divides the phase diagram of the 
$Z_2$ Hamiltonian into several regions. If we were to label these regions in terms of the
$Z_2$ topological index, we would have only two types of regions with $\nu=\pm1$ which
alternate upon crossing every white border in Fig.~\ref{fig:jlambdaforphi}. 

\begin{figure}[t]
    \centering
        \includegraphics[width=1\linewidth]{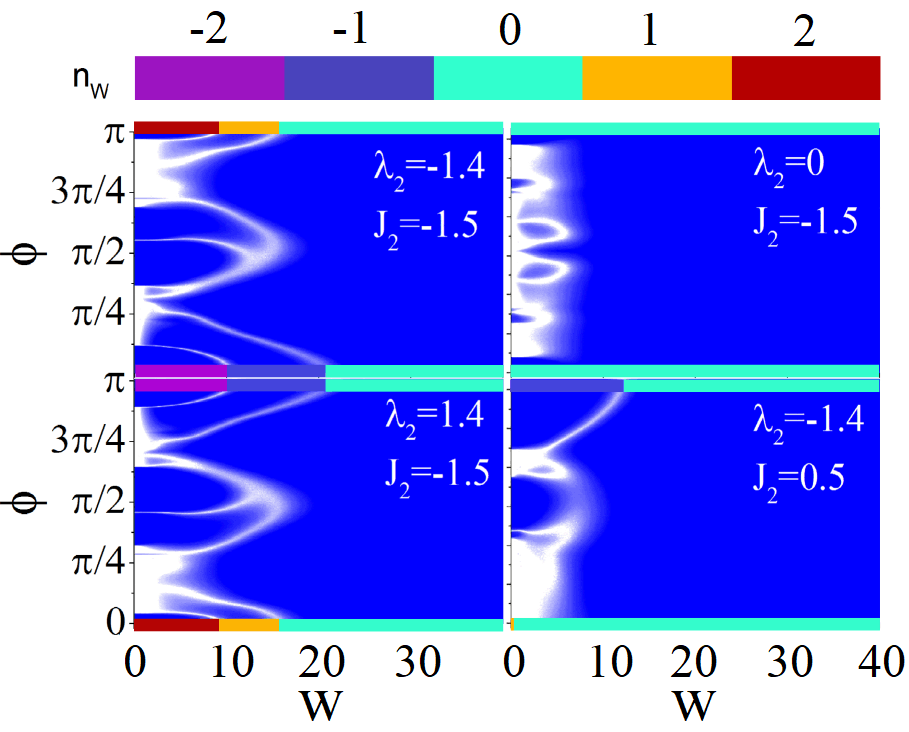}
        %\includegraphics[width=.4\linewidth]{./w-phi-1}
        %\includegraphics[width=.4\linewidth]{./w-phi-2}
        %\includegraphics[width=.4\linewidth]{./w-phi-3}
        %\includegraphics[width=.4\linewidth]{./w-phi-4}
        %\label{fig:mu012}
    \caption{(Color online)Localization length for selected points in the phase diagram as a function of $\phi$ and $W$. 
    The color code of the margin as indicated in the legend corresponds to the winding number of the parent ($\phi=0$)
    Hamiltonian. The common feature of all these figures is that the ultimate fate of the system in strong disorder regime is to end up with a topologically trivial state. 
    Moreover for small $W$, the broadening role of $\phi\ne0$ mod $\pi$ is manifest. 
    }
    \label{fig:/w-phi}
\end{figure}

Now that we have the localization length as a tool at hand that can diagnose the
topological information of the parent Hamiltonian, let us use this tool to better
understand the interplay between $\phi$ and $W$. In Fig.~\ref{fig:/w-phi} in every panel
we pick a set of parameters $(J_2,\lambda_2)$ that in the parent Hamiltonian corresponds
to a definite winding number. In our previous work, we have studied the effect of 
$W$ alone on the winding numbers, and have found that the generic role of $W$ is to
reduce the magnitude of the winding number~\cite{Habibi-resilience}. We have further found
that this reduction of the absolute value of the winding number by disorder happens in one-by-one steps. 
Let us start with the description of the top left panel in Fig.~\ref{fig:/w-phi} which in the
parent Hamiltonian corresponds to the winding number $n_w=-2$. 

Let us first walk along the $W=0$ line (the vertical axis). By increasing $\phi$ from zero to $\pi$, 
gapless regions appear which correspond to the band intervening pattern in Fig.~\ref{fig:energy-band}. 
If we walk along the horizontal line which corresponds to adding disorder $W$ to the parent Hamiltonian~\cite{Habibi-resilience},
the absolute value of the winding numbers starts to reduce one-by-one upon each enhancement of the localization length
of the zero energy states. This gives the color code in the horizontal border which indicates how the winding number of
the parent Hamiltonian changes with the disorder. For strong disorder, it ultimately ends in the $n_w=0$ state. 
For $\phi=\pi$, the story is similar to the $\phi=0$, except that $\phi=\pi$ essentially corresponds to flipping the sing of $\lambda_2$, which will then place it in a phase with winding number $n_w=+2$. 
This symmetry is nicely seen in the lower-left panel which is essentially the mirror image of the top-left panel. 
Now smoothly departing from the regions coded with winding number colors in the borders of every panel, 
one can visit the entire phase diagram. As long as no white line is crossed, the winding number of the
parent Hamiltonian remains the same. This allows us to tile the regions in this figure, with the winding number
$n_w$ of the parent $\phi=0$ mod $\pi$ Hamiltonian. 
In the clean system lower-right panel corresponds to the winding number, $n_w=1$. 
Since this particular point is very close to the borderline of the clean parent Hamiltonian, 
upon introducing disorder (walk along the horizontal line), the winding number quickly becomes zero which
is indicated with a long margin color bar. For $\phi=\pi$, the clean parent Hamiltonian is deep in the 
$n_w=-1$ phase, and therefore the margin color bar corresponding to $n_w=-1$ is longer. Upon increasing $W$, 
this phase is also eventually transformed into the $n_w=0$ phase. The rest of the phase diagram consists in a
dominant region with $n_w=0$ and hence $\nu=+1$. Upon crossing every white border, the $\nu$ alternates its sign, while $n_w$ changes by one. Similar considerations apply to the top-right panel. 
Let us emphasize that in the daughter Hamiltonian with where $\phi\ne 0$ mod $\pi$, the integer $n_w$ 
is not the topological index anymore, nevertheless, it still counts the number of zero-energy states that are left in the ends of the chain. Since these numbers are not topological numbers anymore, 
the corresponding zero modes are not topologically protected. 

\begin{figure}[t]
    \centering
    \includegraphics[width=0.7\linewidth]{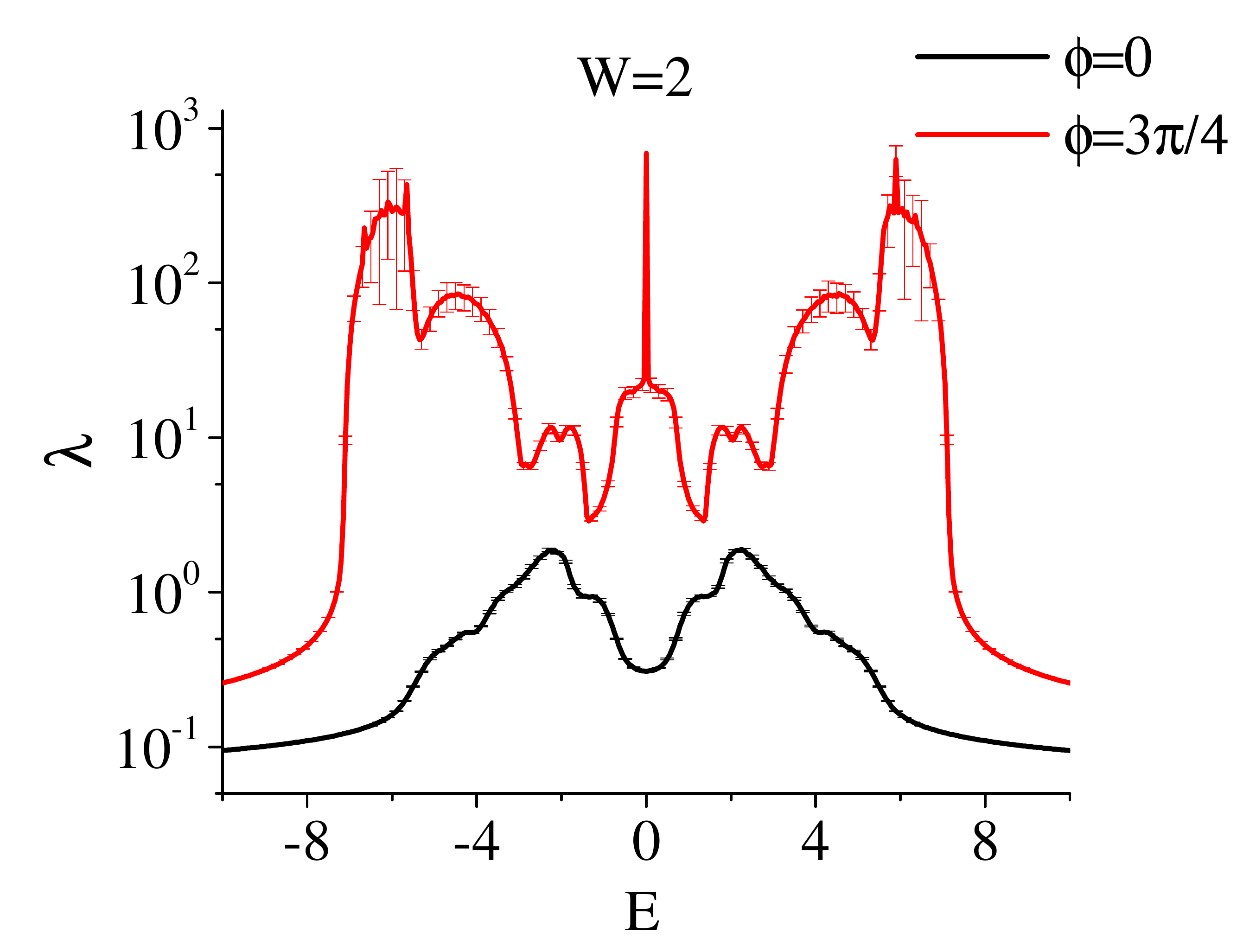}
    \caption{(Color online) Localization length for a disordered system with $\mu=0$, $\lambda_2=-1.4$, $J_2=-1.5$, $W=2$, and with averaging over $200$ configurations. 
    For $\phi=0$ all states are localized and the system is a topological insulator with the winding number $n_w=-2$.  
    For $\phi=3\pi/4$, the localization length of the entire spectrum is remarkably enhanced with respect to the $\phi=0$ case. }
    \label{bright.fig}
\end{figure}

At low-disorder, for some ranges of $\phi$, the zero energy states remain extended
which produce the bright regions in Fig.~\ref{fig:/w-phi}. How does the rest of spectrum look-like
in these regions? In Fig.~\ref{bright.fig} we compare the localization lengths at all energies $E$
for a low-disorder system with $W=2$ corresponding to $\phi=0$ and $\phi=3\pi/4$. In both cases
we have $\lambda=-1.4$ and $J_2=-1.5$.  The $\phi=0$ belongs to the parent Hamiltonian and corresponds to a topological  insulator with $n_w=-2$. The $\phi=3\pi/4$ corresponds to a point deep in the bright region in the top-left panel of Fig.~\ref{fig:/w-phi}. As can be seen in 
Fig.~\ref{bright.fig}, by tuning $\phi$ to $\phi=3\pi/4$ where the intervening between the bands
is achieved, the localization length is markedly enhanced. This is most manifest for $E=0$ extended state.

\subsection{Non-zero chemical potential}
\begin{figure}[t]
    \centering
    \includegraphics[width=\linewidth]{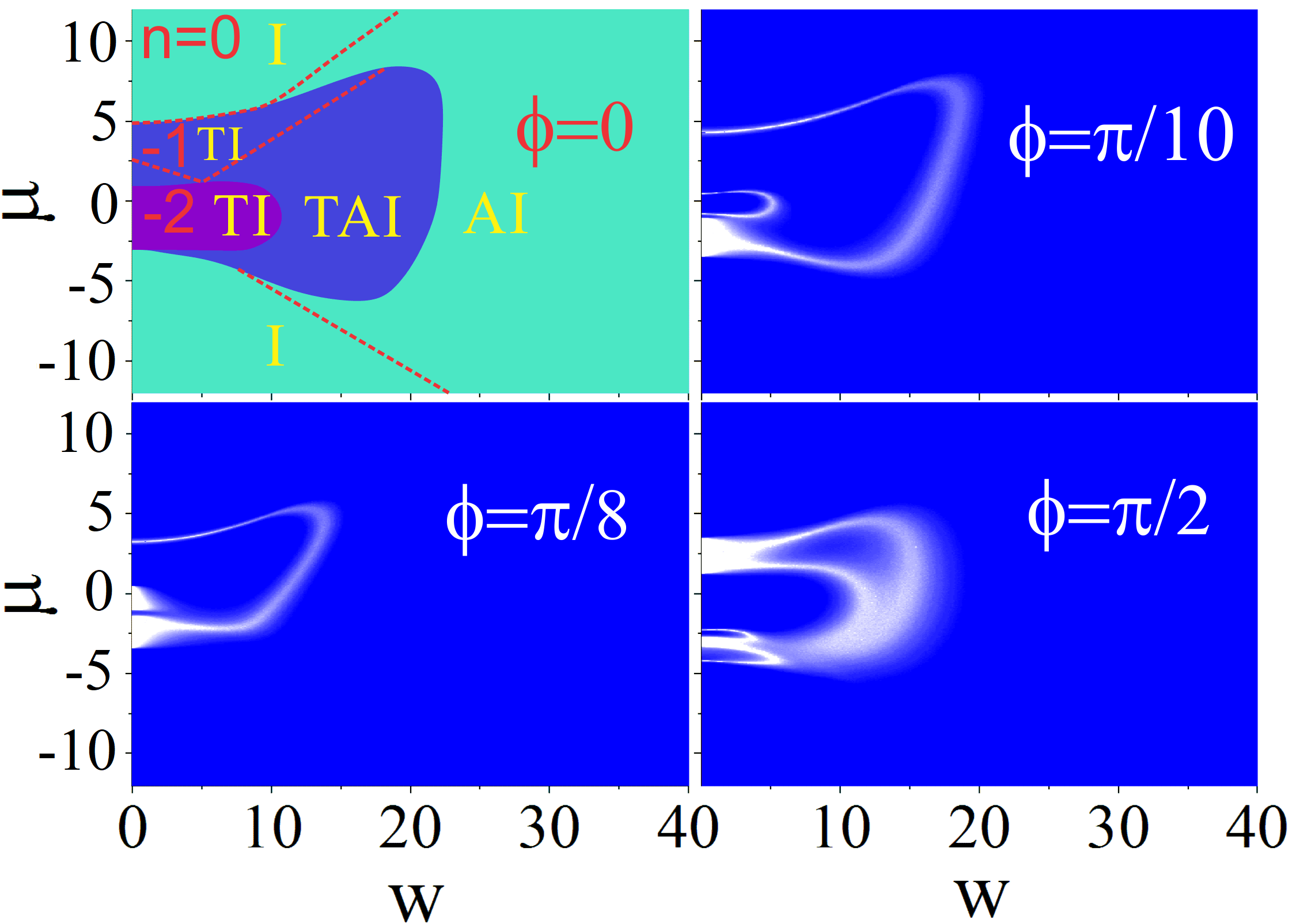} \label{fig:mu017}
    \caption{(Color online)Localization length for different value of $ \phi$ in terms of disorder and constant chemical potential for$\lambda=-1.4,J_2=-1.5$. 
    Increasing $\phi$ from zero to $\frac{\pi}{2}$ partitions the phase diagram into regions with $\nu=\pm1$. 
    Those with same $\nu$ are separated with broad white regions in the low-$W$ regime, while those with different $\nu$ are separated with sharp white borders even in the low-$W$ regime. 
    }
    \label{fig:w-mu-pi}
\end{figure}

So far we have focused on the $\mu=0$ case. It is interesting to study the interplay of non-zero $\mu$ which in the original spin Hamiltonian is equivalent to an applied field along $z$ axis $h\hat k$, Zeeman coupled to spins. 
In Fig.~\ref{fig:w-mu-pi} localization length has been plotted in terms of disorder strength and chemical potential 
for different values of the generalized DM parameter $\phi$. Other parameters are fixed at $J_2=-1.5,~\lambda_2=-1.4$.
In the top-left panel corresponding to $\phi=0$ the system represents the parent Hamiltonian and is characterized by the winding numbers indicated in the figure. Inside the $n_w=-1$ region, the dashed line
represents a line across which the spectral gap of the small $W$ regime is closed by disorder, but the
winding number does not change. The gapless part is actually the topological Anderson insulator, while
the small $W$ part side of the $n_w=-1$ region is a topological insulator. Similar lines exist for $n_w=0$
the region which separates gapped phase from gapless Anderson insulator. However, the Anderson insulator in this case is topologically trivial. 

By moving to the top-right panel where $\phi=\pi/10$, and the topological index is $\nu=\pm1$, as can be seen, the border separating $n_w=-2,0$ of the parent Hamiltonian which corresponds to the same $\nu=+1$ index, is broadened by the DM parameter $\phi$. Upon increasing disorder, this broadening disappears as in the $\mu=0$ examples. 
Upon crossing each bright white line (the divergence of localization length), the magnitude of winding number
changes by $1$. Across the broadened lines it changes by an even number. Therefore the number of (non-protected)
Majorana zero mode pairs in the top-right panel are qualitatively similar to the top-left panel. 
By moving to larger values of $\phi$ in the bottom row, a more complicated pattern can be generated. 
Again regions with different $\nu$ are separated with sharp lines, while those with the same $\nu$
are separated by broad lines in the low-$W$ regime. The broadening is washed away by large $W$. 

\begin{figure}
    \centering
\includegraphics[width=\linewidth]{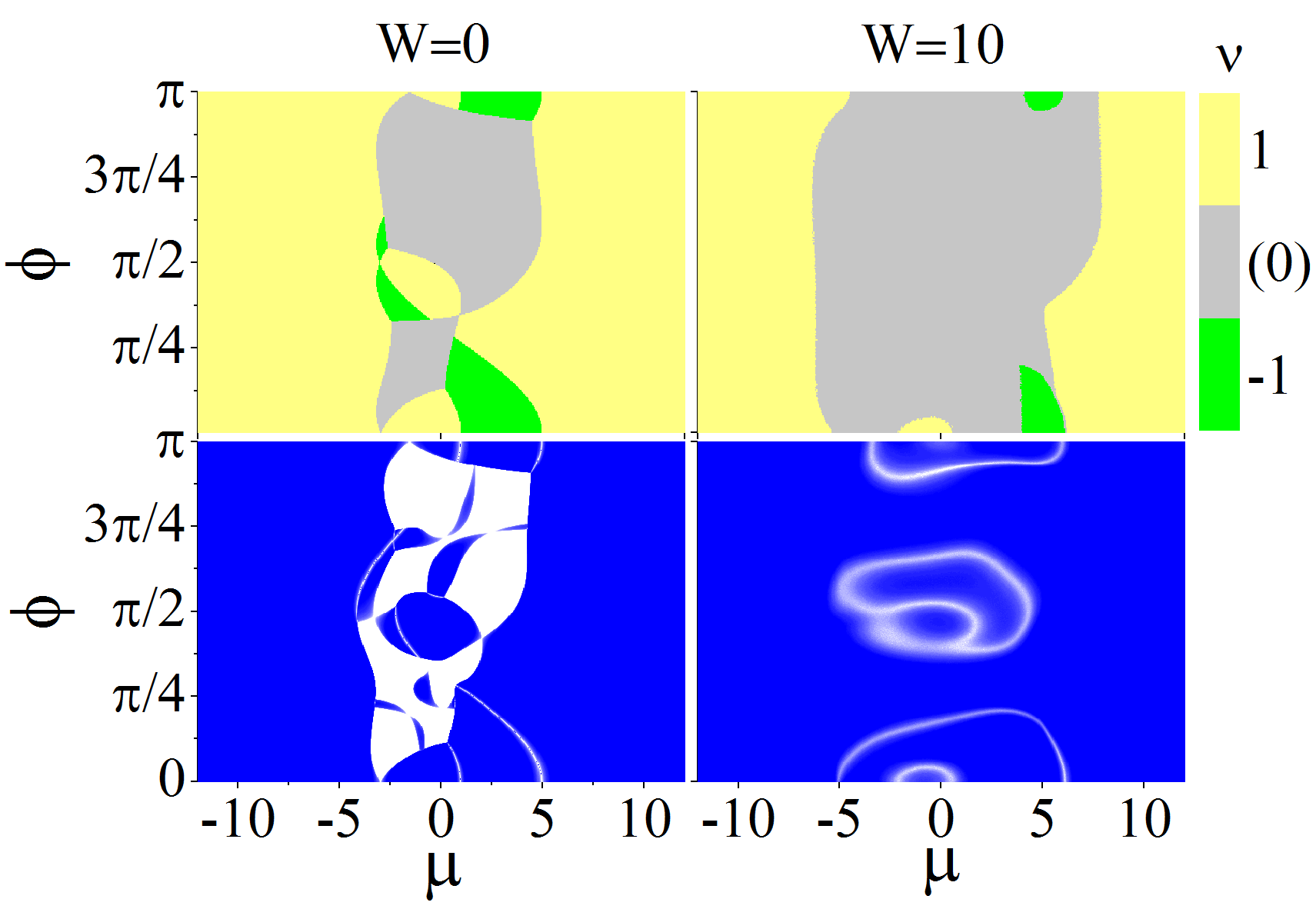} \label{fig:mu021}
    \caption{(Color online) Comparison between the results of Pfaffian analysis and localization length study. 
    Pfaffian is completely blind (gray region) when the system is gapless. However, the localization length, reveals much more information and structure than the topological index determined from Pfaffian.
    }
    \label{fig:mu-phi-W}
\end{figure}

In Fig. \ref{fig:mu-phi-W} we calculate topological phase diagram with localization length and Pfaffian 
analysis in the plane of $\phi$ and $\mu$ for two cases corresponding to $W=0$ (left column) and $W=10$ (right column). 
In the left column for the clean system, we have plotted the phase diagram as determined from
the Pfaffian calculation. The trivial (non-trivial) case, $\nu=+1$ ($\nu=-1$) is denoted by yellow (green). 
The gray region denoted by $\nu=0$ actually means that the Pfaffian is not well defined and it 
signals a gapless situation where the Pfaffian fluctuates between $\pm1$ and eventually averages out to zero.
Physically it means that the system is gapless and Pfaffian can not be defined. 
The bottom panel corresponds to the top panel in each column and is determined from the localization length of the zero energy states. As can be seen by comparing the top and bottom panel in the left column, 
the localization length reveals much more structure than the Pfaffian. Across every sharp line, 
$\nu$ changes, while across the broad while regions $\nu$ stays the same. Similarly for the second column, in presence of strong disorder, according to the Pfaffian, most of the phase
diagram consists in gapless regions (gray) where the Pfaffian is not even well defined. 
But the localization length indicates some structure within the gray region itself. 
Across every white line, the localization length of Majorana fermions inherited from the
parent Hamiltonian at $\phi=0$ critically delocalizes. Despite the system is gapless and
topological index is not even well defined, the localization length
is capable of revealing non-trivial structures.

\section{Conclusion}
In this paper we have studied the topological properties of a generalization of XY model dubbed 2XY with a similarly  
generalized DM coupling in presence of a random transverse field. With the aid of Jordan-Wigner transformation, 
this model can be mapped to 1D higher neighbor hopping Kitaev model with time-reversal symmetry breaking and 
Anderson on-site disorder. The time-reversal symmetry breaking of Jordan-Wigner fermionized model comes from the generalized
DM parameter $\phi$. Any non-zero value of $\phi$ breaks the time-reversal symmetry and reduces the $Z$-valued topological index of the $\phi=0$ (parent)
Hamiltonian to a $Z_2$-valued index $\nu=\pm1$ of the daughter Hamiltonian. 

The result of our previous work~\cite{Habibi-resilience} which deeply roots in the 
bulk-boundary correspondence allows us to focus on the LL of the edge modes at $E=0$ only. 
We used this as a diagnostic tool to reveal information about the TR-restored (parent) model
by looking into the edge modes of the TR-broken (daughter) model. 
The essential lesson from the comparison of the parent and daughter Hamiltonians in this work is that the localization length 
outperforms the topological index in the following respects: (i) In terms of speed, accuracy and efficiency of 
numerical computation, since it only requires the computation for $E=0$ (boundary modes), it will 
be much faster than the computation of the topological index which requires the information of 
the entire spectrum. This simplification roots in the bulk-boundary correspondence. (ii)
In addition to the topological index of the daughter Hamiltonian itself, the LL also contains information 
about the topological index of the parent Hamiltonian. 
Moreover, by tuning $\phi$ due to band-intervening, there appear gapless regions
in the phase diagram where the $Z_2$ topological index is not even well-defined (topological indices
are defined for gapped systems). But the localization length reveals transitions which can not be
captured with the topological index. Therefore the LL of zero modes besides being simpler
to compute contains information beyond the topological index. 

The new insight obtained by the localization length is as follows:
In the daughter Hamiltonian with $\phi\ne 0$, still the Majorana zero modes of the parent Hamiltonian ($\phi=0$)
are localized in the edge, although their edge localization is not protected by winding number anymore. 
By changing various parameters in the Hamiltonian such as the disorder strength, $W$, the Majorana fermions
of the parent Hamiltonian critically delocalize and get drown into the bulk of Anderson localized states. 
In the parent Hamiltonian, this is sensed by a reduction in the magnitude of the winding number. But in the
daughter Hamiltonian where the winding number can not be defined, this is sensed in a different way.
If the divergence in localization length happens on a sharp line, the $Z_2$ index alternates across
the transition line. However, if the transition line is broad -- which happens for low-disorder case -- 
the $Z_2$ index does not even recognize that a pair of Majorana fermions are lost across the transition. 

\section{Acknowledgements}
We wish to acknowledge helpful discussions with Vladimir Kravtsov and Hadi Yarloo. 
We thank Tohid Farajollahpour for insightful discussions on the spin version of the model considered in this work. 

\section{appendix}
\appendix
\section{Transfer Matrix method for Anderson localization}
\label{tm.sec}
\begin{figure}[b]
    \centering
    \includegraphics[width=0.7\linewidth]{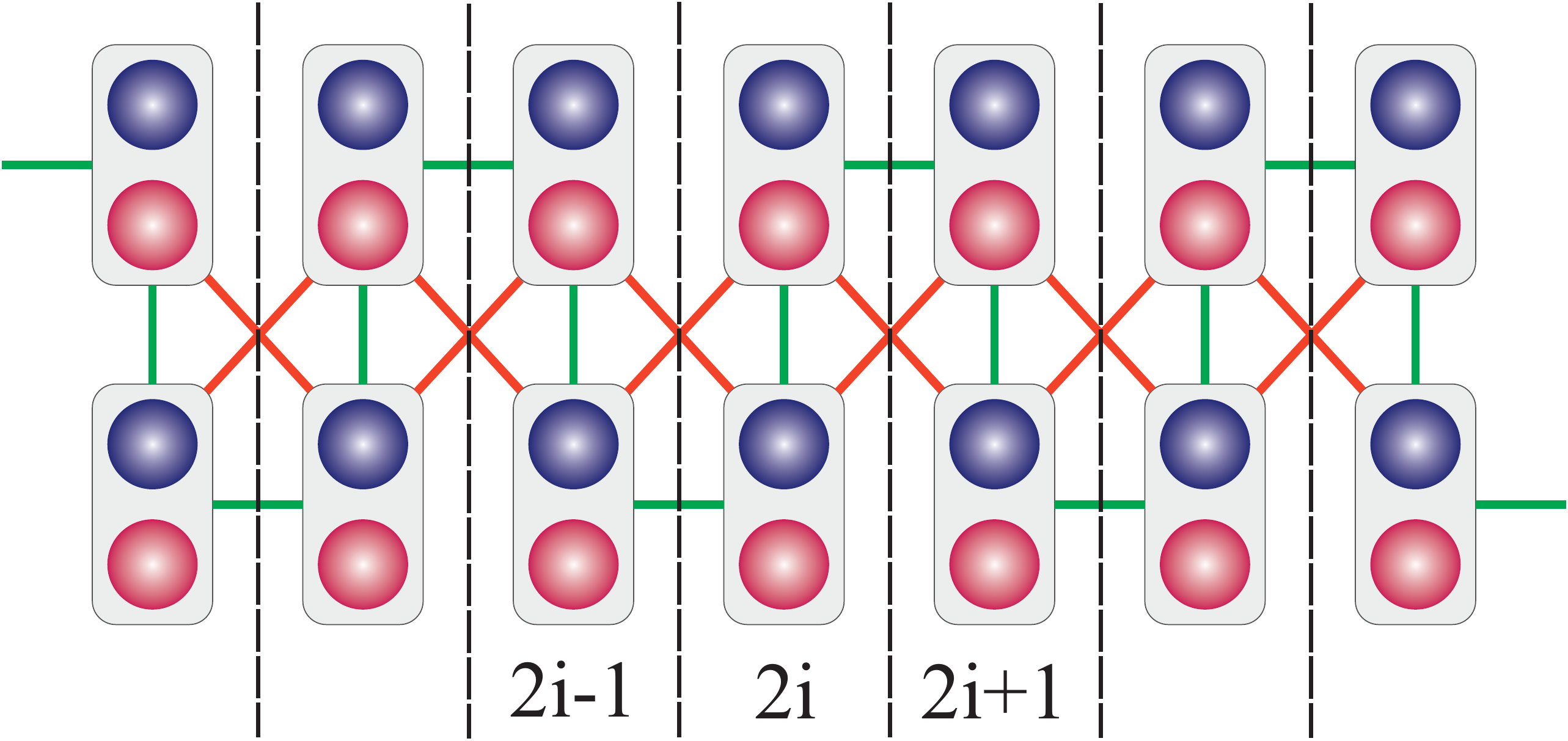}
    \caption{(Color online) Transfer matrix for our Hamiltonian. Each slice contains two atoms. 
    Blue and red circles represent $\psi^e$ and $\psi^h$. Green lines and orange lines show nearest and next nearest neighbor
    couplings.}
    \label{fig:TM}
\end{figure}

To be self-contained, in this appendix we review the transfer matrix method.
This is based on our previous work~\cite{Habibi-resilience}. 
   To calculate the localization length, we can use quasi-one-dimensional Schr\"odinger equation $H{\Psi }_i=E{\Psi }_i$\cite{MacKinnon1,MacKinnon2}. In our model, we need to calculate the localization length for the wave functions in the Nambu space in presence of the generalized DM interaction. 
When we have next nearest neighbor, we are lead to organize the sites into the blocks depicted in Fig.~\ref{fig:TM} such that 
in the newly arranged form, the transfer takes place only between neighboring blocks~\cite{Habibi-resilience}. In this basis every block will have two sites
labeled by indices $1,2$ and the wave function $\Psi_i$ in the Nambu space will be 
$\Psi^T_i=\left(\psi^e_{i,1},\psi^h_{i,1},\psi^e_{i,2},\psi^h_{i,2}\right)$. 
This is effectively a four-channel quasi-one-dimensional problem. 
Within this representation, the wave equation becomes, 
\begin{align}
    t_{i,i-1}^*\vec{\Psi }_{i-1}+H_{i,i}\vec{\Psi }_i+t_{i,i+1}\vec{\Psi }_{i+1}=E \vec{\Psi }_i 
\end{align}
\begin{align}
    \left(
    \begin{array}{c}
        \vec{\Psi }_{i+1} \\
        \vec{\Psi }_{i} 
    \end{array}
    \right) =
    T_{i+1,i}
    \left(
    \begin{array}{c}
        \vec{\Psi }_{i} \\
        \vec{\Psi }_{i-1} 
    \end{array}\right)
\end{align}
where:
\begin{align}
    T_{i+1,i}=\left(
    \begin{array}{cc}
        t^{-1}_{i,i+1} \left(E-H_{i,i}\right) &- t^{-1}_{i,i+1} t_{i,i-1} ^*\\
        1 & 0 \\
    \end{array}
    \right).
\end{align}

As can be seen in the Fig .~\ref{fig:TM} we have two kind of slices labeled by $ 2i $ and $ 2i+1 $, respectively.   
Hopping and onsite matrix for each slice can be written as:

\begin{multline}
    t_{2i,2i+1} =t_{2i+1,2i}^T =\left(
    \begin{array}{cccc}
        0& 0 &J_2 & \lambda _2 \\
        0 & 0 &-\lambda _2 & -J_2^*   \\
        J_2 & \lambda _2 & J_1 & \lambda _1 \\
        -\lambda _2 & -J_2^*  & -\lambda _1 & -J_1^* \\
    \end{array}
    \right),
    \\
    t_{2i+1,2i+2} =t_{2i+2,2i+1}^T=\left(
    \begin{array}{cccc}
        J_1 & \lambda _1 &J_2 & \lambda _2 \\
        -\lambda _1 & -J_1^*   &-\lambda _2 & -J_2^*   \\
        J_2 & \lambda _2 & 0 &0\\
        -\lambda _2 & -J_2^*  &0 & 0\\
    \end{array}
    \right),
    \\
    H_{2i,2i}=\left(
    \begin{array}{cccc}
        \mu+\epsilon _{i,1} & 0 & J_1 & \lambda _1 \\
        0 & -\mu-\epsilon _{i,1} & -\lambda _1 & -J_1^*  \\
        J_1 ^* & -\lambda _1 & \mu+\epsilon _{i,2} & 0 \\
        \lambda _1 & -J_1 & 0 & -\mu-\epsilon _{i,2} \\
    \end{array}
    \right),
    \\
    H_{2i+1,2i+1}=\left(
    \begin{array}{cccc}
        \mu+\epsilon _{i,1} & 0 & J_1 & -\lambda _1 \\
        0 & -\mu-\epsilon _{i,1} & \lambda _1 & -J_1^*  \\
        J_1 ^* & \lambda _1 & \mu+\epsilon _{i,2} & 0 \\
        -\lambda _1 & -J_1 & 0 &-\mu -\epsilon _{i,2} \\
    \end{array}
    \right).
\end{multline}
To calculate the localization length, one needs to conctruct the product of T-matrices as,
\begin{align}
    T_{N,1}=\prod_{i=1,N}{T_{i+1,i}}.
\end{align}
Then the localization length Λ is numerically computed as,
\begin{align}
    \Lambda=\frac{1}{\gamma_{\rm min}}.
\end{align}
where the smallest positive Lyapunov exponent $\gamma_{\rm min}$ is
defined by the eigenvalues {$e^{\gamma_i}$
    ; $i = 1\ldots 8$} of the matrix,
\begin{align}
    \Gamma=  \lim_{N \to \infty}{\left[\prod_{i=N,1}{T^\dagger_{i+1,i}\prod_{i=1,N}{T_{i+1,i}}}\right]^{1/{2N}}}.
\end{align}
Details concerning the numerical method of obtaining
the smallest positive Lyapunov exponent precisely 
are discussed in Ref.~\onlinecite{MacKinnon1,MacKinnon2}.
In our calculation, $ N $ will be chosen large enough to ensure that localization length converges. 

\bibliographystyle{apsrev4-1}
\bibliography{Refs}
\end{document}